# Exploiting Atomic Layer Deposition for Fabricating Sub-10 nm X-ray Lenses

Benedikt Rösner,[1*] Frieder Koch,[1] Florian Döring,[1] Jeroen Bosgra,[1] Vitaliy A. Guzenko,[1] Eugenie Kirk,[1,2] Markus Meyer,[3] Joshua L. Ornelas,[3] Rainer H. Fink,[3] Stefan Stanescu,[4] Sufal Swaraj,[4] Rachid Belkhou,[4] Benjamin Watts,[1] Jörg Raabe,[1] Christian David[1]

[1] Paul Scherrer Institut, 5232 Villigen PSI, Switzerland

[2] Laboratory for Mesoscopic Systems, Department of Materials, ETH Zürich, 8093 Zürich, Switzerland

[3] Department Chemistry and Pharmacy, FAU Erlangen-Nürnberg, Egerlandstr. 3, 91058 Erlangen, Germany

[4] Synchrotron SOLEIL, L'ormes des Merisiers, Saint Aubin BP-48, 91192, Gif-Sur-Yvette Cedex, France

## Abstract:

Moving towards significantly smaller nanostructures, direct structuring techniques such as electron beam lithography approach fundamental limitations in feature size and aspect ratios. Application of nanostructures like diffractive X-ray lenses require feature sizes of below 10 nm to enter a new regime in high resolution X-ray microscopy. As such dimensions are difficult to obtain using conventional electron beam lithography, we pursue a line-doubling approach. We demonstrate that this method yields structure sizes as small as 6.4 nm. X-ray lenses fabricated in this way are tested for their efficiency and microscopic resolution. In addition, the line-doubling technique is successfully extended to a six-fold scheme, where each line in a template structure written by electron beam lithography evolves into six metal lines.

# 1. Introduction

The fabrication of nanostructures with small lateral sizes and high aspect ratios is one of the key challenges in developing highly resolving diffractive X-ray optics with sufficient efficiency. While direct writing of dense nanostructures with electron beam lithography (EBL) becomes extremely challenging below 20 nm feature size [1-6], growth processes can achieve extreme precision, sometimes even down to the atomic level. Utilizing such processes to double the line density of patterns written by EBL enables production of diffractive X-ray lenses, i.e. Fresnel zone plates (FZPs), with the smallest line width below the resolution limit of conventional EBL [1, 7].

The diffraction-limited spot size of a FZP is generally comparable to the outermost zone width [8]. Thus, the achievable minimum dimension of its zones is the limiting factor in terms of resolution in X-ray microscopy. Going beyond diffraction-limited spot sizes of 10 nm in X-ray microscopy, line widths have to be decreased accordingly to 8 nm or below. As fabrication of the required sub-10 nm nanostructures is impossible with conventional EBL, the line-doubling method relaxes the demand of the lithography step to the writing of only every second line. Utilizing atomic layer deposition (ALD) of iridium allows for doubling, or even further multiplying the line density of the template structure [9, 10].

This scheme for enhancing the resolution of lithography has been applied to fabricate FZPs with line widths down to 12 nm [1, 7, 11]. Optimizing this approach further, quadrupling of the line density has been achieved via stacking two line-doubled structures on both sides of a silicon nitride membrane [12]. This approach resulted in 7 nm structures with aspect ratios exceeding 40. However, severe limitations are induced by the alignment precision, making nanostructures written in a single lithography step more appealing for fabrication of X-ray lenses. Especially in the soft X-ray regime, where reasonable efficiency can be achieved with smaller aspect ratios (~ 10), structure sizes below 10 nm may be realized using the line-doubling approach.

In this paper, we report on the fabrication of sub-10 nm metal patterns using line multiplication. We differentiate between two approaches: i) the line-doubling technique established to define patterns with varying periodicities for the exact placement of iridium lines as required for FZPs. In this approach, a template pattern is fabricated by EBL using a negative-tone resist. The template is designed to fill exactly every second gap of the resulting metal nanostructure for an accurate placement of the metal lines. ALD of iridium is then used to deposit the respective nanostructure, and subsequent argon milling and chemical etching can remove the template structure, if required. Using this approach, we have successfully fabricated a set of FZPs with outermost zone widths of 8.8, 8.0, 7.2 and 6.4 nm at aspect ratios between five and ten. ii) A more sophisticated scheme of subsequent ALD, argon milling and etching steps can be employed to achieve multiplication of metal lines by more than a factor of two. We have applied this approach to six-fold template structures with 20 nm thick iridium lines.

## 2. Line-doubling to Fabricate Sub-10 nm X-ray Lenses

### 2.1. Fabrication Procedure

The processing flow chart for line-doubling is illustrated in Figure 1. As first step, a template structure is written by EBL onto an 80 nm thick silicon nitride membrane using a negative tone resist. As a standard, we use hydrogen silsesquioxane (HSQ) which converts to silicon dioxide when exposed to an electron beam. We expose the resist with a Raith EBPG 5000+ lithography tool operated at 100 keV. In order to push the lithographic process to the limit, we choose a beam current of 250 pA and a mid-column aperture size of 200 μm resulting in the smallest beam size available. After e-beam exposure, the resist is developed in Microposit 351 developer mixed with water (1:3), rinsed with water and stored in isopropanol. To avoid collapse of the zones, the samples are dried from supercritical $CO_2$ in a table top critical point dryer (Leica EM CPD 300 Auto). This leads to a sparse pattern on the substrate consisting of every second gap in the final metal nanostructure. In a subsequent step, iridium is deposited onto the template structure using a Picosun R200 ALD system. A typical cycle consists of dosing an iridium precursor (Ir(acac)$_3$) at 370°C and removing the organic part of the precursor with oxygen plasma (50 sccm gas flow, inductively coupled plasma power 2 kW). The thickness of the iridium layer can be controlled by variation of the number of ALD cycles. In this way, the template structures are coated with a uniform iridium layer. However, practical use as lenses for soft X-rays might require the removal of the caps and bottom parts of the iridium coating, and the HSQ template itself, as these parts of the structure act as absorbers. Therefore, the horizontal parts of the iridium pattern is removed by argon milling (Oxford Ionfab 300) in a collimated argon ion beam with an energy of 400 eV under an angle of 2° from the surface normal to minimize the risk of re-deposition. Afterwards, HSQ is etched in hydrofluoric acid (HF) vapour. The remaining nanostructure then consists of free-standing iridium lines. Each fabrication step is documented in detail using a Zeiss Supra 55 VP scanning electron microscope (SEM).

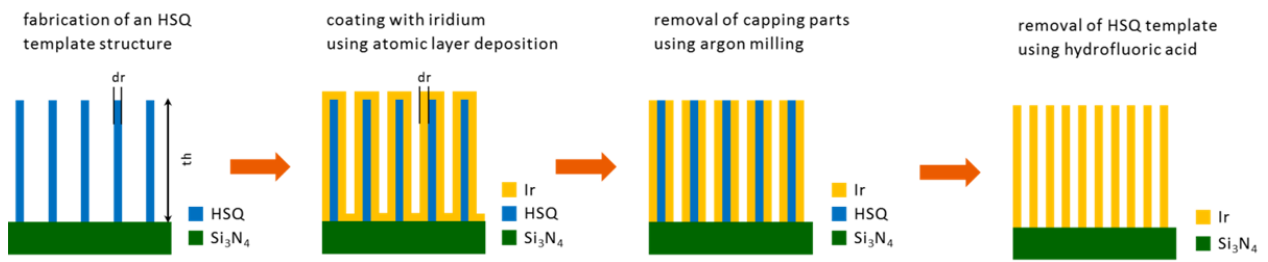

**Figure 1: Flow chart showing the principle of line-doubling. The four steps depict the fabrication of an HSQ template structure with EBL, deposition of an iridium layer with a defined thickness, removal of the iridium cap and bottom utilizing argon milling and removing the template structure with HF vapour (from left to right).**

As the optimum efficiency of line-doubled zone plates depends on the exact line width of both iridium and the HSQ template [13], we prepared a set of 16 zone plates for each zone plate design with systematic variation of exposure dose and ALD cycles. Using four different exposure doses ensures coverage of a range in HSQ thickness. Each set of four zone plates with different exposure doses is then processed separately in the ALD system, allowing variation of the iridium thickness as a function of deposition cycles. As both thicknesses are influenced by variations inherent to processing, the ideal combination of HSQ and iridium thickness has to be determined experimentally by measuring the diffraction efficiency of the zone plates.

## 2.2. Resulting Fresnel Zone Plates

The design parameters for FZP with outermost line widths below 10 nm were chosen to match the experimental conditions at the scanning transmission X-ray microscopes (STXM) at the PolLux and Hermes beamlines at the Swiss Light Source and Synchrotron Soleil [14, 15]. Special attention has been given to the illumination conditions at these beamlines to ensure that the zone plate dimensions match the coherence length and energy bandwidth. The following table shows the parameters for the zone plates fabricated:

Table 1: Design parameters of the fabricated zone plates. Outermost zone widths have been designed below 10 nm. Diameters of 240 μm were chosen for the Hermes beamline and 100 μm for the PolLux beamline. Focal lengths and depths of focus result from diameter, outermost zone width and energy.

| outermost zone width [nm] | diameter [μm] | design energy (E) [eV] | number of zones | focal length [mm] (at design E) | depth of focus [nm] (at design E) |
|---|---|---|---|---|---|
| **8.8** | 240 | 850 | 6818 | 1.45 | 213 |
| **8.0** | 240 | 850 | 7500 | 1.32 | 176 |
| **7.2** | 240 | 850 | 8333 | 1.19 | 142 |
| **6.4** | 240 | 850 | 9375 | 1.05 | 112 |
| **8.8** | 100 | 850 | 2841 | 0.60 | 213 |
| **8.0** | 100 | 850 | 3125 | 0.55 | 176 |

Figure 2 shows scanning electron micrographs of the different steps during the fabrication of a FZP designed with 8.8 nm outermost zone width. The nanostructures have been fabricated as described above. Careful characterization of the nanostructures showed that the desired width of both HSQ and iridium lines is close to their design value of 8.8 nm. In this particular example we measured an HSQ thickness of 8.7 nm and an Ir thickness of 9.2 nm from an average of three linear profiles in each SEM image. The height of the nanostructures was determined from SEM images recorded under an angle of 45° and found to be 70-80 nm, and the structural integrity appears to be high.

As pointed out before, we need to remove part of the iridium layer in order to avoid undesired absorption losses. We used argon ion milling to open the iridium structures, as depicted in Figure 2c. The resulting nanostructure consists of a repetitive sequence of iridium, HSQ, iridium, and a gap. In a fourth step, the HSQ template can be removed in HF vapour to yield the pure iridium structure (Figure 2d). However, numerical simulations with rigorous coupled wave theory [16, 17] indicate that the negative effect of the HSQ template on the diffraction efficiency is negligible in the geometry of our zone plates. We therefore left the HSQ template in the nanostructures to improve the mechanical stability of the zones.

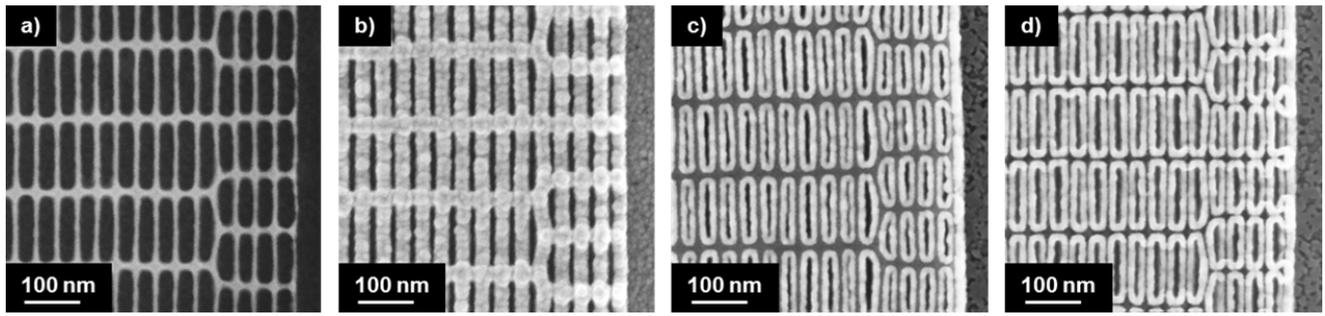

**Figure 2: Resulting structures of an 8.8 nm zone plate. HSQ template (a), after coating with iridium (b), after argon milling (c), and after HF vapour etch (d). The iridium line width is approx. 9 nm, the height 70-80 nm.**

To examine how small the template pattern can be fabricated, a dedicated test pattern with line widths down to 5 nm was designed and processed. The pattern contained lines with decreasing line width and periodicity similar to a FZP, allowing us to determine the minimum line width, which can be exposed and developed in sufficient quality. The minimum applicable dimensions were found to correspond to an FZP pattern with 6-7 nm outermost line width. At smaller dimensions, the HSQ lines were not fully developed, bent or showed high defect density.

Figure 3 shows the template and iridium-coated structures of FZPs with 8.0, 7.2 and 6.4 nm outermost line width. All structures were subsequently opened with argon milling (not shown). SEM images and first efficiency tests with X-rays indicate that the structures down to 7.2 nm are of high structural quality. However, the structural fidelity of the template structure for the FZP with 6.4 nm wide lines indicates that the lithography step is reaching its limit. A first efficiency test of the zone plates with X-rays at 750 eV still yielded a reasonable diffraction efficiency of 1.3%, confirming the overall integrity of the structures. In addition, the height of the structures is reduced to approx. 60 nm (40 nm) for the 7.2 nm (6.4 nm) zone plates.

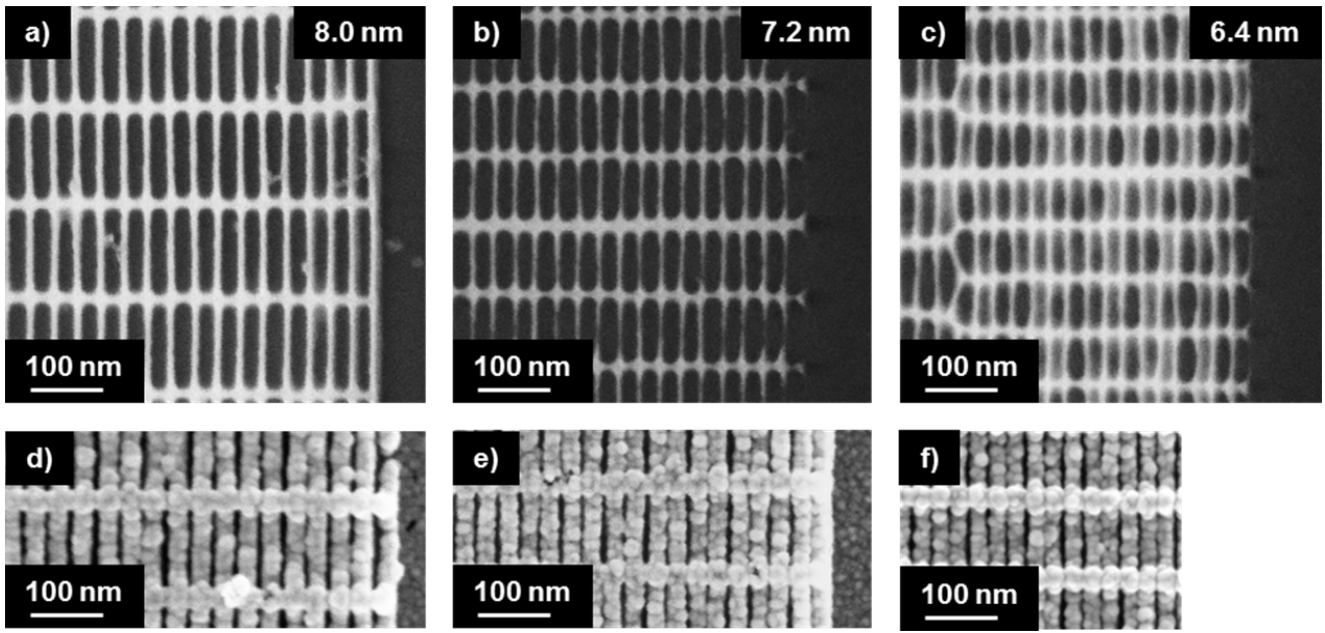

**Figure 3: Sub-10 nm zone plates. HSQ template for 8.0 nm (a), 7.2 nm (b) and 6.4 nm (c). The respective nanostructures coated with iridium are shown in (d-f).**

## 2.3. Efficiency and Resolution Tests

Each of the FZPs has been tested at the PolLux beamline to determine its diffraction efficiency. For this purpose, a 5 μm wide pinhole has been scanned through the focal plane of the zone plate, and the transmitted intensity $I$ has been recorded with a photomultiplier tube. From reference scans without zone plate and supporting $Si_3N_4$ membrane to determine the incoming X-ray intensity $I_0$, the diffraction efficiency can be calculated.

Maximum diffraction efficiencies were found to be 2.6% for 8.8 nm FZPs, 1.8% for 8.0 nm, 1.5% for 7.2 nm, and 1.3% for 6.4 nm at a photon energy of 750 eV. Simulations using rigorous coupled wave analysis [16] yielded 6.4%, 6.0%, 5.2% and 4.2% for ideal, binary gratings with the iridium line widths mentioned above, a height of 80 nm and on 80 nm silicon nitride substrates. Note that these numbers are overestimated in the case of line-doubled zone plates as the efficiencies vary depending on the number of ALD cycles and exposure doses [13]. We thus determined two zone plates for each combination of zone width and diameter, which showed the highest diffraction efficiency for further resolution tests.

Successful resolution tests were performed in the STXM at the Hermes beamline. We used an FZP with a diameter of 240 μm, an outermost zone width of 8.8 nm, and a central stop made of gold with 80 μm diameter and approx. 3 μm thickness. The photon energy was set to 850 eV, and the order-sorting aperture had a diameter of 30 μm. This results in a focal length of 1.45 mm, a depth of focus of 213 nm, and a calculated diffraction-limited spot size of 10.7 nm [8].

For an initial, preliminary test of the zone plate's resolution under these conditions, we scanned a test sample with 10 nm iridium lines and spaces, which was fabricated in a similar way as the FZPs. Figure 4 shows a highly resolved scan of this test sample with a step size of 1 nm. The micrograph shows well-separated lines and spaces with 20 nm periodicity. The visibility (~1%) is much lower than the maximum possible visibility of 60 nm thick iridium structures (40%) as the modulation transfer function of the zone plate goes towards zero when approaching the resolution limit.

The recorded test structure indicates that our zone plates are indeed capable of producing extremely small X-ray spots which can potentially resolve feature sizes below 10 nm. Previous studies using FZPs with outermost zone widths of 17 nm and 12.5 nm achieved resolutions of 10 nm at 700 eV [18] and 9 nm at 1.2 keV [1], respectively. In view of these values, we expect to achieve sub-10 nm resolution in X-ray microscopy in ongoing tests which will provide fully coherent illumination of our zone plates.

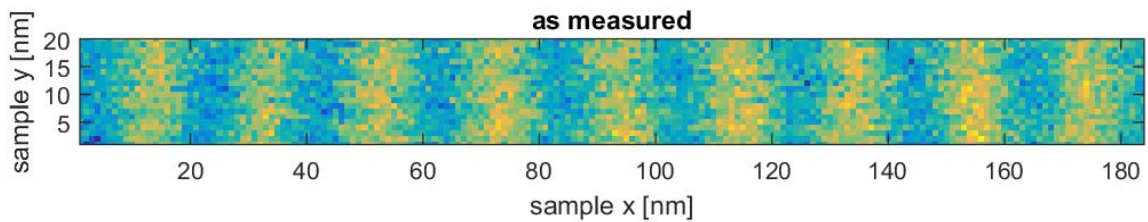

**Figure 4: Scanning transmission X-ray micrograph of an iridium test sample with 10 nm lines and spaces, taken at the Hermes beamline. The photon energy is 850 eV, the step size is 1 nm. The visibility of the lines and spaces is approx. 1%.**

# 3. Six-fold Multiplication of Iridium Lines

## 3.1. Fabrication Scheme

Taking the line-multiplication approach further, the free-standing metal lines obtained can be further coated with different materials by ALD. For this purpose, the initial template structure has to be designed with wider lines to obtain an iridium template with even sparser lines than in the line-doubling approach (see Figure 5). The subsequent steps consist of ALD of aluminium oxide ($Al_2O_3$) from a trimethylaluminum precursor at 300°C, argon milling removing the caps again, deposition of iridium and again an argon milling step. As a final step, the $Al_2O_3$ lines are removed by plasma etching using a $SF_6/O_2$ plasma having flows of 25/2.5 sccm, RF power of 150 W and a pressure of 2.7 Pa.

The approach to six-fold iridium lines can only be used for nanostructures that are designed with constant periodicities. For the fabrication of FZPs, where each line has to be placed at a defined radius, its position has to be controlled by the template written in the EBL step. In contrast, the six-fold scheme will always yield three metal lines with equal distance defined by the intermediate $Al_2O_3$ deposition step. This scheme is therefore suitable especially for gratings, resolution test samples, or ring structures with constant pitch.

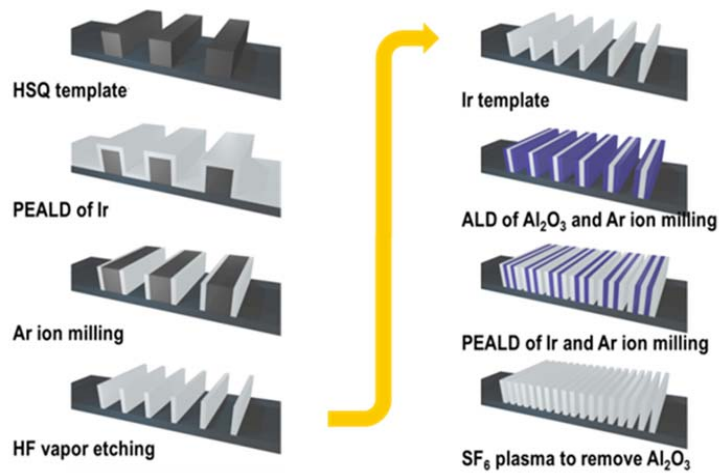

**Figure 5: Principle of six-fold multiplication of iridium lines. The left column of the flow chart corresponds to the steps described in Figure 1, with the difference that the initial template is programmed with wider lines. The resulting line-doubled iridium structure is then taken as template for further processing (right column): ALD of $Al_2O_3$ and argon milling, plasma-enhanced ALD of iridium and argon milling, and a final $Al_2O_3$ etch.**

## 3.2. Resulting Nanostructures

To go beyond the feature density of the line-doubling approach, we investigated metal lines which were six-fold multiplied from an HSQ template. As a proof-of-principle experiment, we designed a nanostructure with 20 nm wide iridium lines. SEM images of the different processing steps are shown in Figure 6. As initial template, an array of 150 nm thick HSQ cuboids was fabricated and covered with 20 nm of iridium (a). After argon milling and removal of HSQ in HF vapour, this results in a structure consisting of every third iridium line of the final pattern (b), which acts simultaneously as template for further processing. Subsequently, $Al_2O_3$ is deposited via ALD on the iridium lines, the caps are removed by argon milling, and a new layer of iridium is deposited (c). Note that the outer side walls are not fully perpendicular to the substrate. Most likely, the observed side wall slope is an artefact related to inevitable faceting of the lines during Ar ion milling [19]. After ion milling of the final iridium layer, the artefact due to the faceting is more apparent (d). A difference in height of the inner and outer iridium lines can be observed. From SEM measurements we find that the outer lines are 65 nm high, whereas the inner lines are 115 nm high. This difference is explained from the geometry of the initial cuboid. Faceting during ion milling of the first iridium layer occurs only on the outside of the iridium lines. The inside of the initial iridium lines is always occupied either by iridium itself, or, during over-etching of iridium, by the HSQ template. The final etching step with $SF_6$ then yields a metal pattern (e), where a set of six lines can be identified, three for each edge of the initial HSQ template structure.

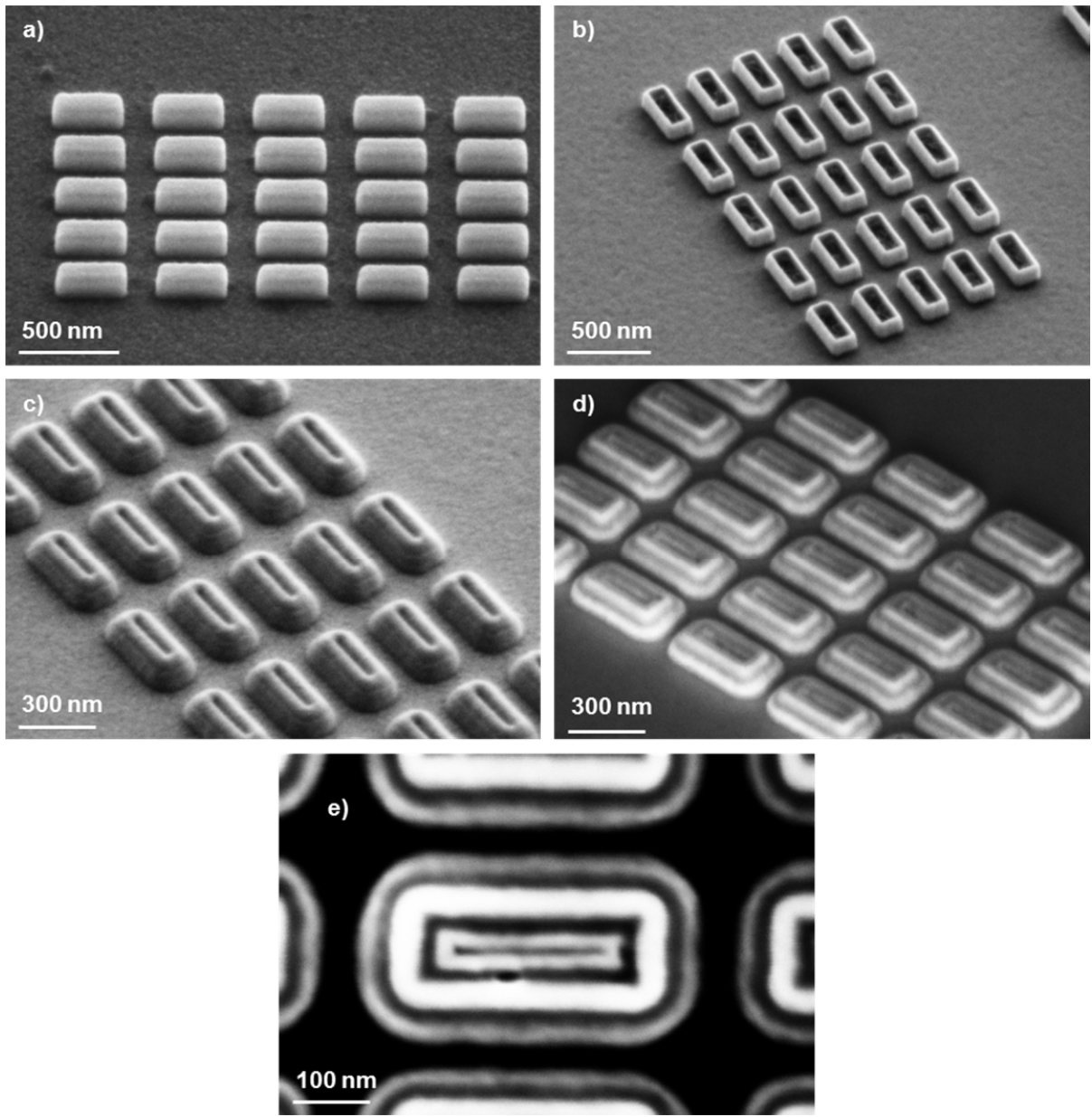

**Figure 6:** Scanning electron microscopy images of six-fold multiplied iridium lines. a) template consisting of 150 nm thick HSQ structures with 300 nm x 100 nm lateral dimensions, and covered with 20 nm iridium, b) 20 nm iridium lines after Ar milling and HF etch, c) 6-folded iridium structure after ALD of 20 nm $Al_2O_3$, Ar milling, and ALD of 20 nm iridium, d) after Ar milling of the final iridium layer, e) after $SF_6$ plasma etch to remove the $Al_2O_3$.

## 4. Conclusions

We have demonstrated a line multiplication approach to fabricate metal nanostructures with line widths in the sub-10 nm regime. The line-doubling approach is successfully applied to the fabrication of FZPs with diffraction-limited resolution approaching 10 nm and below. As an additional advantage, the placement of each metal line can be fully controlled by the design of the template structure, making this approach especially appealing for focusing optics, such as FZPs, but also off-axis zone plates [20, 21], or one-dimensional gratings with a line focus.

We expect that the achievement to fabricate FZPs with feature sizes well below 10 nm will have a major impact on X-ray microscopy. In state-of-the art X-ray microscopes, routinely achieved resolution is in the order of 30-50 nm, i.e. typical spot sizes are significantly larger. We resolved feature sizes in the order of 10 nm using the FZP with the widest outermost zone width of 8.8 nm fabricated in the scope of this study. In view of newly developed diffraction-limited X-ray sources, highly coherent synchrotron and free-electron laser radiation will be available [22]. These developments will allow to use our FZPs with even smaller zones and thus lead to a notable enhancement of resolution in X-ray microscopy.

Furthermore, we have shown that it is possible to extend the line-doubling approach to a line multiplication with six lines in the final nanostructure with respect to its template. However, this method shows that the structural quality starts to suffer from the intermediate processing steps, limiting practical use of the multiplication scheme. Moreover, the exact thickness of each layer has to be carefully adjusted in three ALD steps. Nevertheless, six-fold multiplication of metal lines offers a perspective to fabricate high-resolution gratings in aspect ratios that line-doubling cannot provide. This possibility has to be evaluated in future work.

## Acknowledgements

This work received funding from the EU-H2020 Research and Innovation Programme, No. 654360 NFFA-Europe, under the Marie Skłodowska-Curie grant agreement No. 701647, and the EU's Seventh Framework Programme (FP7/2007-2013) (290605, COFUND: PSIFELLOW). MM, JOL and RHF acknowledge financial support by the German Federal Minister of Education and Research (grant 05K16WED). The PolLux end station was financed by the German Ministerium für Bildung und Forschung (BMBF) through contracts 05KS4WE1/6 and 05KS7WE1.

This publication is dedicated to the memory of our dear colleague Jeroen Bosgra who passed away unexpectedly. The co-authors are deeply saddened that this promising young researcher's life has come to an end and that Jeroen was not able to carry on with his research. It is an honour to publish some of his last contributions in the field.

## References

[1] J. Vila-Comamala, K. Jefimovs, J. Raabe, T. Pilvi, R.H. Fink, M. Senoner, A. Maaßdorf, M. Ritala, C. David, Advanced Thin Film Technology for Ultrahigh Resolution X-ray Microscopy, Ultramicroscopy 109 (2009) 1360-1364.
[2] W. Chao, B.D. Harteneck, J.A. Liddle, E.H. Anderson, D.T. Attwood, Soft X-ray microscopy at a spatial resolution better than 15 nm, Nature 435(7046) (2005) 1210-1213.
[3] W. Chao, J. Kim, S. Rekawa, P. Fischer, E.H. Anderson, Demonstration of 12 nm Resolution Fresnel Zone Plate Lens based Soft X-ray Microscopy, Opt. Express 17(20) (2009) 17669-17677.
[4] J. Reinspach, M. Lindblom, M. Bertilson, O. von Hofsten, H.M. Hertz, A. Holmberg, 13 nm high-efficiency nickel-germanium soft x-ray zone plates, J. Vac. Sci. Technol. B 29(1) (2011) 011012.
[5] J. Reinspach, F. Uhlén, H.M. Hertz, A. Holmberg, Twelve nanometer half-pitch W–Cr–HSQ trilayer process for soft x-ray tungsten zone plates, J. Vac. Sci. Technol. B 29(6) (2011) 06FG02.
[6] S. Rehbein, P. Guttmann, S. Werner, G. Schneider, Characterization of the resolving power and contrast transfer function of a transmission X-ray microscope with partially coherent illumination, Opt. Express 20(6) (2012) 5830-5839.
[7] K. Jefimovs, J. Vila-Comamala, T. Pilvi, J. Raabe, M. Ritala, C. David, Zone-Doubling Technique to Produce Ultrahigh-Resolution X-Ray Optics, Phys. Rev. Lett. 99(26) (2007) 264801.
[8] M. Born, E. Wolf, Elements of the Theory of Diffraction, Principles of Optics, Pergamon Press, Oxford, 1970, p. 464.
[9] L.F. Johnson, K.A. Ingersoll, Generation of surface gratings with periods < 1000 Å, Appl. Phys. Lett. 38(7) (1981) 532-534.
[10] D.C. Flanders, N.N. Efremow, Generation of <50 nm period gratings using edge defined techniques, J. Vac. Sci. Technol. B 1(4) (1983) 1105-1108.
[11] J. Vila-Comamala, S. Gorelick, E. Färm, C.M. Kewish, A. Diaz, R. Barrett, V.A. Guzenko, M. Ritala, C. David, Ultra-high resolution zone-doubled diffractive X-ray optics for the multi-keV regime, Opt. Express 19(1) (2011) 175-184.

[12] I. Mohacsi, I. Vartiainen, B. Rösner, M. Guizar-Sicairos, V.A. Guzenko, I. McNulty, R. Winarski, M.V. Holt, C. David, Interlaced zone plate optics for hard X-ray imaging in the 10 nm range, Sci. Rep. 7 (2017) 43624.
[13] F. Marschall, J. Vila-Comamala, V.A. Guzenko, C. David, Systematic efficiency study of line-doubled zone plates, Microelectron. Eng. 177 (2017) 25-29.
[14] J. Raabe, G. Tzetkov, U. Flechsig, M. Böge, A. Jaggi, B. Sarafimov, M.G.C. Vernooij, T. Huthwelker, H. Ade, D. Kilcoyne, T. Tyliszczak, R.H. Fink, C. Quitmann, PolLux: A New Facility for Soft X-ray Spectromicroscopy at the Swiss Light Source, Rev. Sci. Instrum. 79 (2008) 113704.
[15] R. Belkhou, S. Stanescu, S. Swaraj, A. Besson, M. Ledoux, M. Hajlaoui, D. Dalle, HERMES: a soft X-ray beamline dedicated to X-ray microscopy, Journal of Synchrotron Radiation 22(4) (2015) 968-979.
[16] G. Schneider, Zone plates with high efficiency in high orders of diffraction described by dynamical theory, Appl. Phys. Lett. 71(16) (1997) 2242-2244.
[17] S. Rehbein, A. Lyon, R. Leung, M. Feser, G. Schneider, Near field stacking of zone plates for reduction of their effective zone period, Opt. Express 23(9) (2015) 11063-11072.
[18] W. Chao, P. Fischer, T. Tyliszczak, S. Rekawa, E. Anderson, P. Naulleau, Real space soft x-ray imaging at 10 nm spatial resolution, Opt. Express 20(9) (2012) 9777-9783.
[19] R.E. Lee, Microfabrication by ion-beam etching, J. Vac. Sci. Technol. 16(2) (1979) 164-170.
[20] F. Marschall, D. McNally, V.A. Guzenko, B. Rösner, M. Dantz, X. Lu, L. Nue, V. Strocov, T. Schmitt, C. David, Zone plates as imaging analyzers for resonant inelastic x-ray scattering, Opt. Express 25(14) (2017) 15624-15634.
[21] F. Marschall, Z. Yin, J. Rehanek, M. Beye, F. Döring, K. Kubiček, D. Raiser, S.T. Veedu, J. Buck, A. Rothkirch, B. Rösner, V.A. Guzenko, J. Viefhaus, C. David, S. Techert, Transmission zone plates as analyzers for efficient parallel 2D RIXS-mapping, Scientific Reports 7(1) (2017) 8849.
[22] M. Eriksson, J.F. van der Veen, C. Quitmann, Diffraction-limited storage rings - a window to the science of tomorrow, J. Synchrotron Rad. 21 (2014) 837-842.